\documentclass[11pt,letterpaper]{article}
\usepackage[T1]{fontenc}
\usepackage[utf8]{inputenc}
\usepackage{lmodern}
\usepackage[margin=1in]{geometry}
\usepackage{microtype}
\usepackage{authblk}
\usepackage{amsthm,amsmath,amsfonts,amssymb,mathtools}
\usepackage[authoryear,round]{natbib}
\usepackage{graphicx,bm,booktabs,enumitem,bbm,float,longtable,multirow}
\usepackage{xcolor}
\usepackage{placeins}
\usepackage{xspace}
\usepackage[colorlinks,citecolor=blue,urlcolor=blue,linkcolor=red]{hyperref}
\graphicspath{{image/}{./}}
\theoremstyle{plain}

\theoremstyle{definition}

\newcommand{\Nor}{\mbox{N}}
\newcommand{\logN}{\mbox{log-N}}
\newcommand{\IG}{\mbox{inv-Ga}}

\newcommand{\Be}{\text{Be}}

\newcommand{\Bern}{\text{Bern}}
\newcommand{\NB}{\text{NB}}

\newcommand{\bY}{\bm{Y}}

\newcommand{\bth}{{\bm \Theta}}
\newcommand{\blam}{{\bm \lambda}}
\newcommand{\bLam}{{\bm \Lambda}}
\newcommand{\bSig}{{\bm \Sigma}}
\newcommand{\beeta}{{\bm \eta}}

\newcommand{\Exp}{\mbox{E}}

\newcommand{\Cov}{\mbox{Cov}}

\newcommand{\iid}{\stackrel{iid}{\sim}}

\def\ind{\mathbbm{1}}

\def\T{{\mathrm{\scriptscriptstyle T} }}

\newcommand{\suppcomp}{A\xspace}
\newcommand{\suppaddsim}{B\xspace}
\newcommand{\suppsens}{C\xspace}
\newcommand{\suppaddreal}{D\xspace}

\title{Bayesian Nonparametric Longitudinal Factor Model for Microbiome Counts}
\author[1]{Shuangjie Zhang\thanks{Corresponding author: \href{mailto:shuangjie.zhang@austin.utexas.edu}{shuangjie.zhang@austin.utexas.edu}.}}
\author[2]{Seoyeon Ok}
\author[3]{Arul Jayaraman}
\author[4]{Kirsten Landrock}
\author[5]{Ivan Ivanov}
\author[4]{Robert S. Chapkin}
\author[1]{Yang Ni}
\affil[1]{Department of Statistics and Data Science, The University of Texas at Austin}
\affil[2]{Department of Statistics, Texas A\&M University}
\affil[3]{Artie McFerrin Department of Chemical Engineering, Texas A\&M University}
\affil[4]{Department of Nutrition, Texas A\&M University}
\affil[5]{Department of Veterinary Physiology \& Pharmacology, Texas A\&M University}
\date{}
\begin{document}
\maketitle
\begin{abstract}
Longitudinal microbiome studies provide important insights into the temporal evolution of microbial abundances and dynamic cross-feature dependence. Modeling longitudinal microbiome data is challenging because microbiome counts are highly heterogeneous and zero-inflated, and, moreover, repeated measurements induce within-subject temporal dependence and possibly time-varying cross-feature covariance. To address these challenges, we develop a flexible Bayesian longitudinal factor model using a rounded log-normal latent representation. The mean abundance process begins with a Dirichlet process mixture baseline, which accommodates sparse and multimodal abundance distributions, and subsequently evolves through an autoregressive process. Autoregressive coefficients differ across prespecified biological groups, which are defined by genotypes in our application. A separate autoregressive factor process captures shared cross-feature dependence. 
The model therefore distinguishes subject--feature mean evolution from shared dependence and yields a closed-form, time-indexed covariance. Simulation studies evaluate dynamic covariance recovery under matched and misspecified data-generating mechanisms. In the longitudinal mouse-gut microbiome application, LODI identifies highly persistent feature-specific abundance trajectories and shared latent factors that distinguish the three genotype groups over time. At week 15, the dominant factor follows the observed WT $<$ ACK/G $<$ HACK/G tumor-severity ordering, with its strongest contributions arising from OTUs classified within Clostridiaceae, \textit{Clostridium}, and RF39.
\end{abstract}

\section{Introduction}\label{sec:intro}

High-throughput sequencing technologies in microbial ecology generate high-dimensional multivariate counts that characterize microbial communities such as the gut, skin, and oral microbiome. The data are typically represented as a table of operational taxonomic units (OTUs) or amplicon sequence variants \citep{li2015microbiome}. Increasingly, these samples are collected repeatedly from the same subjects to study microbial development, resilience, disease progression, and responses to interventions \citep{gerber2015longitudinal,lyu2023methodological,creus2025methodology}. One important inferential objective is to determine how cross-feature dependence evolves over time.

Generally, microbiome count data present several well-known statistical difficulties. Sequencing depth varies substantially across samples, excess zeros arise from both biological absence and limited detection, nonzero counts are highly overdispersed, and the number of taxa may be comparable to or larger than the number of subjects \citep{paulson2013differential,li2015microbiome}. Longitudinal sampling introduces two additional challenges. First, repeated samples from the same subject are temporally dependent, violating the independence assumption used by many cross-sectional procedures. Second, temporal variation contains at least two conceptually different components: a subject-feature mean trajectory describing how an individual taxon's abundance changes, and a shared cross-feature process describing covariance variation among taxa. Without an explicit separation, feature-specific mean evolution can be absorbed into the estimated covariance and create misleading dependence patterns.

\begin{figure}[h]
\centering
\begin{tabular}{@{}c@{}}
\includegraphics[width=0.98\textwidth]{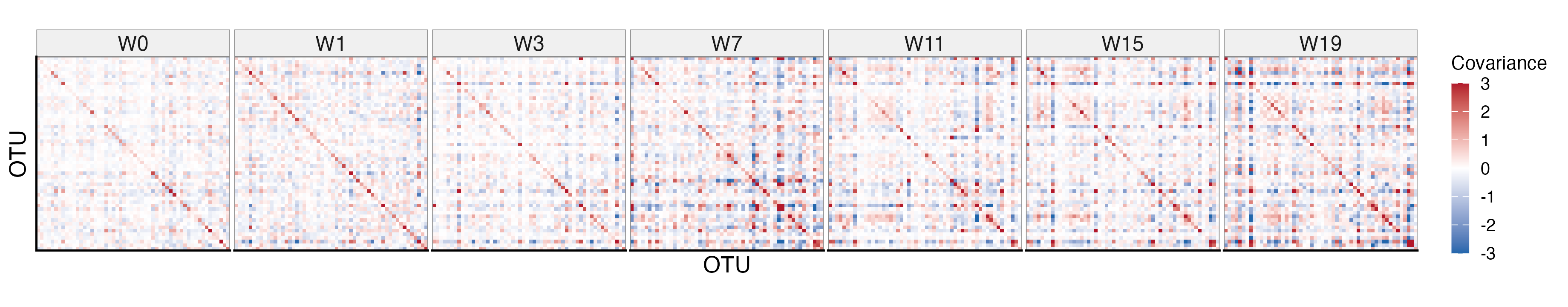}\\[2pt]
(a) Empirical OTU covariance at each time point.\\[6pt]
\includegraphics[width=0.62\textwidth]{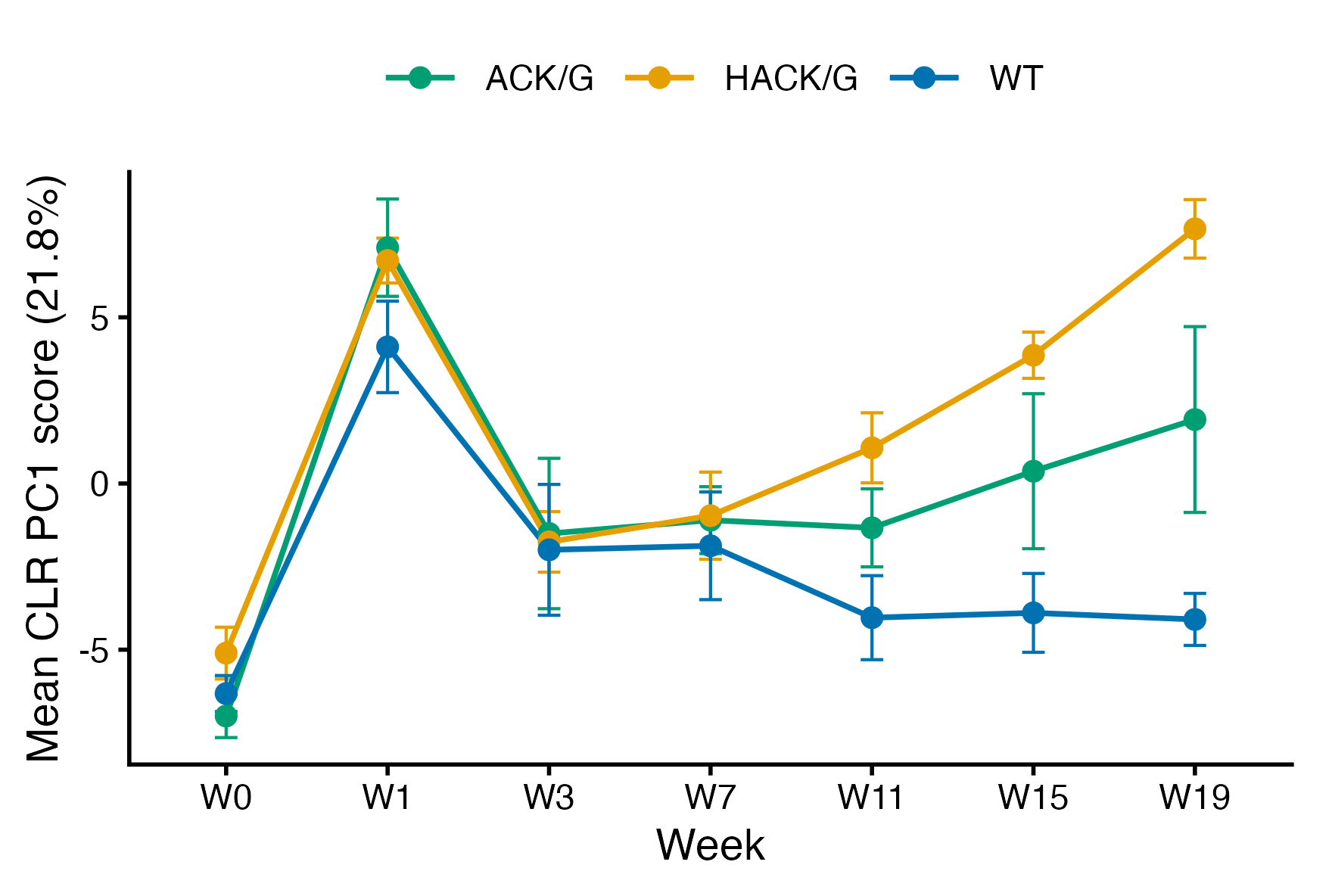}\\[2pt]
(b) Genotype-specific CLR principal-component trajectories.
\end{tabular}
\caption{[Real Data] In (a), each heatmap shows the empirical covariance matrix among the 54 CLR-transformed OTUs, calculated across the 20 mice at the indicated time point. In (b), points and error bars represent genotype-specific means and standard errors of the first principal component scores obtained from a common PCA of all 140 CLR-transformed subject-time profiles.}
\label{fig:intro-data-motivation}
\end{figure}

These challenges arise directly in our motivating longitudinal mouse gut microbiome study. Specifically, the colon cancer dataset contains 20 mice from three genotype groups, ACK/G ($n=4$), HACK/G ($n=10$) harboring $Apc^{S580/+}$ and $Kras^{G12D/+}$ mutations, with/without two alleles of $AhR$ ($AhR^{f/f}$, abbreviated as H), and WT (wild-type control; $n=6$), measured at weeks 0, 1, 3, 7, 11, 15, and 19 \citep{han2021loss} for 54 OTUs. Across the resulting 140 longitudinal samples, 18.9\% of the observed counts are zero overall, with OTU-specific zero rates ranging from 0\% to 92.1\%. Figure~\ref{fig:intro-data-motivation}(a) shows the empirical covariance matrix of the 54 CLR-transformed OTUs across the 20 mice at each time point. Broad covariance patterns are shared across time points, while the magnitudes and signs of the covariances between specific OTU pairs vary visibly over time. 
Figure~\ref{fig:intro-data-motivation}(b) presents genotype-specific mean scores and standard errors for the first principal component, obtained from a common PCA of all 140 CLR-transformed profiles. This component explains 21.8\% of the total variation, and its trajectories exhibit increasing separation among the genotype groups after week 7, particularly between HACK/G and WT. These patterns support a longitudinal model that allows feature interactions and count abundance to differ across known biological groups and evolve over time.

Mixture models address some of the aforementioned challenges by representing samples or latent parameters as arising from multiple subpopulations. A prominent example is the Dirichlet-multinomial mixture (DMM) \citep{holmes2012dirichlet}, which models each microbial composition using one of a finite number of Dirichlet-multinomial components. It accommodates unequal library sizes and overdispersion while clustering samples into latent metacommunity types. \citet{gibbons2017dynamic} further extended DMM to a longitudinal subject-specific version applied to human gut microbiome time series data. Beyond finite-mixture formulations, Bayesian nonparametric priors have been used in microbiome analysis for several related but distinct purposes. \citet{ren2017bayesian} constructed dependent normalized random measures with latent factors to jointly model microbial distributions and propagate uncertainty. \citet{shuler2021bayesian} used dependent DP priors for covariate-indexed distributions of zero-inflation probabilities and differential-abundance effects in multivariate count regression. More recently, Sp-BGFM \citep{zhang2025sparse} modeled multi-domain microbiome count vectors through a DP mixture of rounded multivariate log-normal kernels, coupling flexible baseline abundance distributions with a group factor structure for within- and cross-domain covariance. This formulation demonstrates how DP mixtures can accommodate zeros and substantial between-sample heterogeneity while avoiding an overly restrictive parametric baseline. 

For modeling high-dimensional microbial dependence, existing methods largely fall into two categories. The first category estimates a sparse covariance or precision matrix after transforming the counts. SparCC \citep{friedman2012inferring} estimates pairwise log ratio covariance between latent log-transformed abundances by approximating Pearson correlations. CCLasso \citep{fang2015cclasso} similarly uses log ratio transformed compositions but estimates a sparse and positive-definite log-basis covariance matrix through $\ell_1$-penalized least squares. SPIEC-EASI \citep{kurtz2015sparse} applies a centered log-ratio transformation followed by sparse precision-matrix estimation, and Zi-LN \citep{prost2021zero} instead uses a CLR transformation and a latent Gaussian model to estimate the sparse precision matrix. See also REBACCA \citep{ban2015investigating}, COAT \citep{cao2019large} and phyloBCG \citep{chung2022phylogenetically}. The second category assumes a latent low-dimensional representation of the high-dimensional covariance matrix. For example, ZI-MLN \citep{zhang2023AOAS} is a latent factor model that combines a rounded multivariate log-normal kernel with shrinkage priors on the loadings.
Related low-rank and covariance methods have also been applied to transformed compositional and multi-omics profiles, including MOFA \citep{argelaguet2018multi}, robust covariance estimation for compositional data \citep{he2021robust}, and zero-inflated logistic-normal principal components analysis \citep{zeng2023zero}. These methods provide useful cross-sectional estimates, but for longitudinal microbiome data, fitting them independently at each time point ignores temporal dependence, whereas pooling all time points together ignores changes in covariance over time.

Methods developed specifically for longitudinal data have largely focused on modeling abundance trajectories. MC-TIMME uses negative binomial observation models with piecewise exponential mean dynamics and clusters OTUs by their responses through a Dirichlet-process mixture \citep{gerber2012inferring}. Multivariate sparse functional principal-component analysis instead represents subject-level multi-omics trajectories through smooth basis expansions \citep{jiang2022bayesian}. These approaches characterize temporal changes in individual features but are not designed for time-varying covariance estimation. Related dynamic count models include hierarchical negative binomial state-space models \citep{chapados2014effective} and gamma-process dynamic Poisson factor analysis \citep{acharya2015nonparametric}. Although both approaches borrow information over time, their parametric count distributions are insufficiently flexible to accommodate the excessive zeros and heterogeneous abundance distributions common in microbiome data. MEFISTO introduces Gaussian-process factors for continuous multi-omics measurements \citep{velten2022identifying}, but analysis of zero-inflated microbial counts generally requires a transformation or a separate observation model. Thus, an important gap remains for a model that jointly provides a flexible count distribution, a longitudinal mean process, and a dynamic factor process for time-varying covariance.

We address this gap with LODI, a new Bayesian longitudinal factor model for microbiome counts that explicitly decomposes the temporal signal into a subject-feature mean process and a shared cross-feature factor process. We first use a rounded multivariate log-normal representation from \citet{canale2011bayesian} to handle the discreteness of microbiome counts. Then, we let the initial value of the temporal mean process follow a Dirichlet-process mixture \citep{sethuraman1994constructive}, allowing heterogeneous and multimodal baseline abundances, and propagate it through a group-specific AR(1) process. Next, we use a time-invariant loading matrix paired with an AR(1) factor process, resulting in a dynamic covariance. The abundance and factor components are distinguished through their different hierarchical and temporal structures. 
LODI provides subject- and feature-specific abundance trajectories whose autocorrelation varies across groups (genotypes in our application). In parallel, the shared factor process yields within-time and lagged cross-time covariance matrices, allowing feature-pair dependence to be tracked over time.

The remainder of the paper is organized as follows. Section~\ref{sec:method} presents the model, prior specification, and posterior computation. Section~\ref{sec:simulation} evaluates LODI in simulation studies, and Section~\ref{sec:real-data} analyzes longitudinal mouse-gut microbiome data. Section~\ref{sec:discussion} concludes with directions for future work.

\section{Model}\label{sec:method}

\subsection{Rounded log-normal observation model}

Let $Y_{ijt}$ denote the observed count for feature $j=1,\ldots,P$ from subject $i=1,\ldots,N$ at time $t=1,\ldots,T$. Each subject belongs to a known group $g(i)\in\{1,\ldots,G\}$, such as a genotype. Following the rounded-kernel construction of \citet{canale2011bayesian}, we introduce a log-normal positive latent variable 
\begin{equation}
Y_{ijt}^{\star}
\sim\logN(\mu_{ijt}, \sigma_t^2),
\label{eq:latent-observation}
\end{equation}
where  $\mu_{ijt}$ is the  location parameter and $\sigma_t^2$ is the variance parameter. The observed count is obtained through $Y_{ijt}=\lfloor Y_{ijt}^{\star}\rfloor$.
Then, for every nonnegative integer $y$,
\begin{equation}
\begin{aligned}
\Pr\!\left\{Y_{ijt}=y\mid\mu_{ijt},\sigma_t^2\right\}
&=\int_y^{y+1}
f_{\logN}\!\left\{y^\star\mid\mu_{ijt},\sigma_t^2\right\}\,dy^\star
\\
&=\Phi\!\left\{\log(y+1)\mid\mu_{ijt},\sigma_t^2\right\}
-\Phi\!\left\{\log y\mid\mu_{ijt},\sigma_t^2\right\}.
\end{aligned}
\label{eq:prob-count}
\end{equation}
Here, $f_{\logN}$ denotes the log-normal density, and $\Phi(\cdot\mid m,v)$ is the normal distribution function with mean $m$ and variance $v$. 

\subsection{Longitudinal mean process}
After specifying the rounded kernel, we decompose its location parameter as
\begin{equation}
\mu_{ijt}=s_{it}+\pi_{ijt}+\blam_j^{\T}\beeta_{it}.
\label{eq:mean-decomposition}
\end{equation}
The subject-specific term $s_{it}$ is shared by all features measured from subject $i$ at time $t$ and absorbs multiplicative differences in sequencing depth or total sample abundance. In particular,
\begin{equation}
\Exp\!\left\{Y_{ijt}^{\star}\mid s_{it},\pi_{ijt},\beeta_{it}\right\}
=\exp\!\left\{s_{it}+\pi_{ijt}+\blam_j^{\T}\beeta_{it}+\frac{\sigma_t^2}{2}\right\},
\label{eq:mean-logN}
\end{equation}
and hence $\exp\{s_{it}\}$ scales every feature in the same subject-time sample by the same amount. Thus, $s_{it}$ is a latent sample-scale factor, which accounts for sample-scale differences, including sequencing depth. From \eqref{eq:mean-logN}, the conditional likelihood alone does not identify $s_{it}$ and $\pi_{ijt}$ individually. For any scalar $d$, the transformation
$
s_{it}\longmapsto s_{it}+d,\ \pi_{ijt}\longmapsto\pi_{ijt}-d,
$
leaves $\mu_{ijt}$ unchanged. In the hierarchical model, we anchor this translation invariance by putting a prior with fixed mean 0, which serves as a soft sum-to-zero constraint.
The process $\pi_{ijt}$ describes subject- and feature-specific abundance after normalization by $s_{it}$. Because it enters the model additively on the log scale, $\pi_{ijt}$ corresponds to normalized baseline relative abundance. Equivalently, $\exp\{\pi_{ijt}-\pi_{iht}\}$ gives the baseline abundance ratio between features $j$ and $h$. The final term in \eqref{eq:mean-decomposition}, $\blam_j^{\T}\beeta_{it}$, captures lower-dimensional variation shared across features and is responsible for the non-diagonal covariance. We will discuss it in detail in Section~\ref{sec:Lambda}.

\subsection{Longitudinal Dirichlet-Process Mixture}

Microbial abundances are often sparse, skewed, and multimodal across subjects and features. A common strategy for excess zeros is a zero-inflated model that combines a zero-generating mechanism with a parametric model for abundance. Examples include the zero-inflated Gaussian model for differential-abundance analysis \citep{paulson2013differential}, the zero-inflated log-normal model for microbial association networks \citep{prost2021zero}, and the Bayesian zero-inflated Dirichlet-multinomial regression model for multivariate compositional counts \citep{koslovsky2023bayesian}. These formulations can distinguish structural absence from sampling variation, but they require a separate zero-generating mechanism and parametric assumptions that identify it from the abundance component. Moreover, an observation attributed to the structural-zero component does not directly inform the positive-abundance density. In contrast, the rounded log-normal kernel in \eqref{eq:prob-count} assigns positive probability to zero through interval censoring. Consequently, zero and nonzero counts inform a common latent abundance distribution. To accommodate remaining multimodality without imposing a single parametric baseline distribution, we place a DP mixture (DPM) prior on the initial value $\pi_{ij1}$ of the log-normalized abundance process $\{\pi_{ijt}\}$. Specifically, using the stick-breaking representation \citep{sethuraman1994constructive}, $
w_\ell=z_\ell\prod_{h<\ell}(1-z_h),\ z_\ell\iid\Be(1,c)
$, we assume
\begin{align*}
\pi_{ij1}&\sim\sum_{\ell=1}^{\infty}w_\ell\,
\Nor(\xi_\ell,\tau_\ell^2),
\\
\pi_{ijt+1}&=\phi_{g(i)}\pi_{ijt}+v_{ijt+1},\qquad
v_{ijt+1}\sim\Nor(0,V).
\end{align*}
Here, $\phi_g\in(-1,1)$ is the first-order autocorrelation parameter in AR(1) for group $g$, and $V$ is the innovation variance. The group-specific AR(1) process propagates the DPM-distributed initial value  $\pi_{ij1}$ through time, providing the needed flexibility without over-complicating the model with a DPM at each time point. We use the following conditionally conjugate priors
$
\xi_\ell\iid\Nor(\xi_0,\kappa_0^2)$, $\tau_\ell^2\iid\IG(a_\tau,b_\tau),\ \phi_g\iid\Nor(\mu_\phi,\sigma_\phi^2)\ind\{-1<\phi_g<1\},
\ V\sim\IG(a_V,b_V).
$ Throughout, $\IG(a,b)$ denotes an inverse-gamma distribution with shape $a$ and rate $b$.

\subsection{Longitudinal factor process}\label{sec:Lambda}

Recall from \eqref{eq:mean-logN} that $\beeta_{it}\in\mathbb{R}^K$ is the factor process for subject $i$, and $\blam_j\in\mathbb{R}^K$ is the loading vector for feature $j$. Let $\bLam=(\blam_1,\ldots,\blam_P)^{\T}$. We assume that the factors follow an AR(1) process:
\begin{equation}
  \begin{aligned}
\beeta_{i1}&\sim\Nor(\bm{0},I_K),
\label{eq:eta-initial}\\
\beeta_{it+1}&=\rho\beeta_{it}+\bm{u}_{it+1},
\qquad \bm{u}_{it+1}\sim\Nor(\bm{0},I_K),
\end{aligned}
\end{equation}
with autocorrelation parameter
$
\rho\sim\Nor(\mu_\rho,\sigma_\rho^2)\ind\{-1<\rho<1\}.
$
Since the likelihood function depends on the loadings and factors only through their product, and for any nonsingular $K\times K$ matrix $H$,
$
\bLam\beeta_{it}
=\left(\bLam H\right)\left\{H^{-1}\beeta_{it}\right\}
$,
the loadings and factors are not separately identifiable without additional restrictions. Fixing both the initial and innovation covariance matrices of the factors at $I_K$ fixes their scale and restricts this invariance to orthogonal transformations. To remove the remaining rotation, permutation, and sign ambiguities, we constrain the first $K$ rows of $\bLam$ to be lower triangular with positive diagonal:
\[
\lambda_{jk}=0\quad\text{for }k>j,
\qquad
\lambda_{jj}>0\quad\text{for }j=1,\ldots,K.
\]
Provided that the leading $K\times K$ block has full rank, the triangular restriction fixes the factor orientation and ordering, while the positive diagonal fixes the signs \citep{geweke1996measuring,lopes2004bayesian}. We place independent $\Nor(0,1)$ priors on the free loading entries. Finally, we assume
$
\sigma_t^2\iid\IG(a_\sigma,b_\sigma)
$
for conjugacy.

Let $\widetilde{\bY}_{it}^{\star}=\log({\bY}_{it}^{\star})=\{\widetilde Y_{i1t}^{\star},\ldots,\widetilde Y_{ijt}^{\star},\ldots,\widetilde Y_{iPt}^{\star}\}^{\T}$. After integrating out the shared factor score, \eqref{eq:eta-initial} implies that the within-time covariance of the latent log-abundance vector is
\begin{equation}
\bSig_t:=
\Cov\!\left\{\widetilde{\bY}_{it}^{\star}\mid s_{it},\bm{\pi}_{it}\right\}
=\frac{1-\rho^{2t}}{1-\rho^2}\bLam\bLam^{\T}+\sigma_t^2I_P.
\label{eq:within-time-covariance}
\end{equation}
For a positive lag $\delta\geq1$, the corresponding cross-time covariance is
\begin{equation}
\Cov\!\left\{\widetilde{\bY}_{it}^{\star},
\widetilde{\bY}_{it+\delta}^{\star}\mid
s_{it},\bm{\pi}_{it},s_{it+\delta},\bm{\pi}_{it+\delta}\right\}
=\rho^\delta\frac{1-\rho^{2t}}{1-\rho^2}\bLam\bLam^{\T}.
\label{eq:cross-time-covariance}
\end{equation}
These latent covariance matrices determine whether two observed counts are independent. In particular, conditional on $s_{it}$ and $\bm{\pi}_{it}$, $\widetilde{\bY}_{it}^{\star}$ is multivariate normal. Thus, for $j\neq j'$, $\Sigma_{jj',t}=0$ implies that $\widetilde Y_{ijt}^{\star}$ and $\widetilde Y_{ij't}^{\star}$ are independent. Because the count transformation is applied separately to each coordinate, their transformed counts are also independent, and hence
\[
\Sigma_{jj',t}=0
\quad\Longrightarrow\quad
\Cov\!\left\{Y_{ijt},Y_{ij't}\mid s_{it},\bm{\pi}_{it}\right\}=0.
\]
The same implication applies across time when the corresponding entry of \eqref{eq:cross-time-covariance} is zero. 

Our formulation uses a common loading matrix $\bLam$ across time so that the same identified factor directions describe dependence throughout the longitudinal study. A more general specification could replace it by $\bLam_t$, giving within-time covariance $\{(1-\rho^{2t})/(1-\rho^2)\}\bLam_t\bLam_t^{\T}+\sigma_t^2I_P$ and cross-time covariance $\rho^\delta\{(1-\rho^{2t})/(1-\rho^2)\}\bLam_t\bLam_{t+\delta}^{\T}$. Such a model permits pair-specific changes in covariance beyond the common temporal scaling in \eqref{eq:within-time-covariance}. However, estimating a separate $P\times K$ loading matrix at each time point is difficult when the number of subjects observed per time point is small relative to the number of features, as in microbiome studies with $N\ll P$. The time-invariant loading matrix therefore pools information across time and provides a more stable covariance estimate, whereas time-varying loadings would require additional temporal smoothing or shrinkage.

\subsection{Posterior computation}

Let $\bth$ collect all unknown quantities. For an observed count $y_{ijt}$, we introduce an augmented latent Gaussian variable restricted to $[\log y_{ijt},\log\{y_{ijt}+1\})$, with lower endpoint $-\infty$ when $y_{ijt}=0$. The augmented posterior is proportional to
\begin{align}
&\prod_{i=1}^N\prod_{j=1}^P\prod_{t=1}^T
\Nor\!\left[\widetilde Y_{ijt}^{\star}\mid
s_{it}+\pi_{ijt}+\blam_j^{\T}\beeta_{it},\sigma_t^2\right]
\nonumber\\[-2mm]
&\hspace{28mm}\times
\ind\!\left[\log y_{ijt}\leq\widetilde Y_{ijt}^{\star}<\log\{y_{ijt}+1\}\right]
\,p(\bth).
\label{eq:augmented-posterior}
\end{align}
Both longitudinal trajectories, $\{\pi_{ijt}\}_{t=1}^T$ and $\{\beeta_{it}\}_{t=1}^T$, are sampled jointly over time using forward-filtering backward-sampling (FFBS). One MCMC iteration proceeds as follows.

\begin{enumerate}[label=\arabic*.]
    \item Sample each $\widetilde Y_{ijt}^{\star}$ from its normal full conditional truncated to $[\log y_{ijt},\log\{y_{ijt}+1\})$, using $\log(0)=-\infty$.

    \item Sample the latent sample-scale term $s_{it}$. Under $s_{it}\sim\Nor(m_s,v_s)$,
    \begin{align}
    s_{it}&\mid\cdot\sim\Nor\{m_{s,it},C_{s,t}\},\nonumber\\
    C_{s,t}=\left\{\frac{P}{\sigma_t^2}+\frac{1}{v_s}\right\}^{-1},
    m_{s,it}&=C_{s,t}\left[
    \frac{m_s}{v_s}+
    \frac{1}{\sigma_t^2}\sum_{j=1}^P
    \left\{\widetilde Y_{ijt}^{\star}-\pi_{ijt}
    -\blam_j^{\T}\beeta_{it}\right\}\right].\nonumber
    \end{align}

    \item Sample the trajectory $\{\pi_{ijt}\}_{t=1}^T$ by FFBS, and update the mixture atoms, stick-breaking weights, group-specific coefficients $\phi_g$, and innovation variance $V$.

    \item Sample the free elements of $\bLam$ subject to its triangular and positive-diagonal restrictions. Update each trajectory $\{\beeta_{it}\}_{t=1}^T$ by FFBS, center the factor scores across subjects at every time point, and then update $\rho$.

    \item Independently for each time point, sample
    \begin{align*}
    \sigma_t^2\mid\cdot\sim\IG\left[
    a_\sigma+\frac{NP}{2},\,
    b_\sigma+\frac{1}{2}\sum_{i=1}^N\sum_{j=1}^P
    \left\{\widetilde Y_{ijt}^{\star}-s_{it}-\pi_{ijt}
    -\blam_j^{\T}\beeta_{it}\right\}^2
    \right].
    \end{align*}
\end{enumerate}
Complete full-conditional derivations are provided in Supplementary Materials Section~\suppcomp.
The implemented sampler uses a finite truncation with a large number $L$ of mixture components.  Both the simulation and data analyses use $L=30$. A component diagnostic assessing the adequacy of this truncation level for the real-data fit is provided in Supplementary Materials Section~\suppaddreal. 
The sampler is implemented in Rcpp and RcppArmadillo for computational efficiency. An R package is available on \url{https://shuang-jie.github.io/LODI}.

\section{Simulation}\label{sec:simulation}

\subsection{Setup and evaluation}

We evaluate LODI's performance with simulations. We first test whether LODI identifies factors and recovers the true covariance under matched-model and ZINB-misspecified conditions. Then, we use real-data features in Section~\ref{sec:real-data} to generate a simulation that mimics real, challenging data. In all simulations, we generate longitudinal count data $\bm{Y} \in \mathbb{Z}_{\ge 0}^{N \times T \times P}$ and measure recovery of the covariance $\bSig_t$ implied by the truth using
$
\text{RMSE}_t=\sqrt{\sum_{j,j'}(\widehat{\bSig}_{jj',t}-\bSig^{\mathrm{true}}_{jj',t})^2/{P^2}}.
$
We report RMSE averaged across time within each replicate and then across the independent replicates available for each simulation setting.

We compare LODI to three existing covariance estimators for microbiome data:
(i) SP-BGFM (per-time)~\citep{zhang2025sparse}, a Sparse Bayesian Group Factor Model fit separately at each time point;
(ii) SP-BGFM (multi-domain), a joint fit that treats the time points as separate data domains without an ordered temporal process for the mean; and
(iii) MEFISTO~\citep{velten2022identifying}, a factor model with a Gaussian process on longitudinal covariates.
SP-BGFM is fitted directly to the observed counts, whereas MEFISTO uses $\log(Y+0.5)$ as its input. 

We fix
$(\xi_0,\kappa_0^2,a_\tau,b_\tau,\mu_\phi,\sigma_\phi^2,a_V,b_V,c)
=(0,25,2,1,0,1,2,1,1)$.
Thus, the mixture-location prior has standard deviation $5$, and the DP concentration parameter is fixed at $c=1$. We test the sensitivity of these parameters in Supplementary Materials Section~\suppsens, which shows our method is reasonably robust within the tested ranges.

\subsection{Simulation results}

\underline{Simulation 1} generates counts directly from the LODI model to evaluate mean-process and covariance recovery under matched conditions and to assess sensitivity to the fitted number of factors. The data-generating design uses $N=20$, $T=7$, $P=50$, $K^{\rm tr}=4$, and $G=3$. The baseline $\pi_{ij1}$ has three mixture components with locations $\{-5,3,8\}$, common variance $0.5$, and weights $\{0.4,0.3,0.3\}$. We set the true abundance-process AR(1) coefficients at $\phi_1^{\rm tr}=-0.8$, $\phi_2^{\rm tr}=0.9$, and
$\phi_3^{\rm tr}=0.7$ for the three groups, respectively, with innovation
variance $V^{\rm tr}=0.2$. These values represent negative temporal dependence
in the first group and positive temporal dependence in the other two
groups. The factor process has $\rho^{\rm tr}=0.7$, initial covariance $I_K$, and innovation covariance $I_K$. The log-scale variance $\sigma_t^{2,\rm tr}$ is drawn independently from $\operatorname{Unif}(0,1)$. Finally, counts are generated as $y_{ijt}=\lfloor\exp\{\widetilde Y_{ijt}^{\star}\}\rfloor$. We fit all methods with $K=K^{\rm tr}=4$. For the proposed LODI, we additionally demonstrate that it is robust with respect to the overspecification of the number of factors $K\in\{6,8,10\}$ (see below and Supplementary Materials Section~\suppsens).

\begin{figure}[h]
\centering
\begin{tabular}{cc}
  \includegraphics[width=0.46\textwidth]{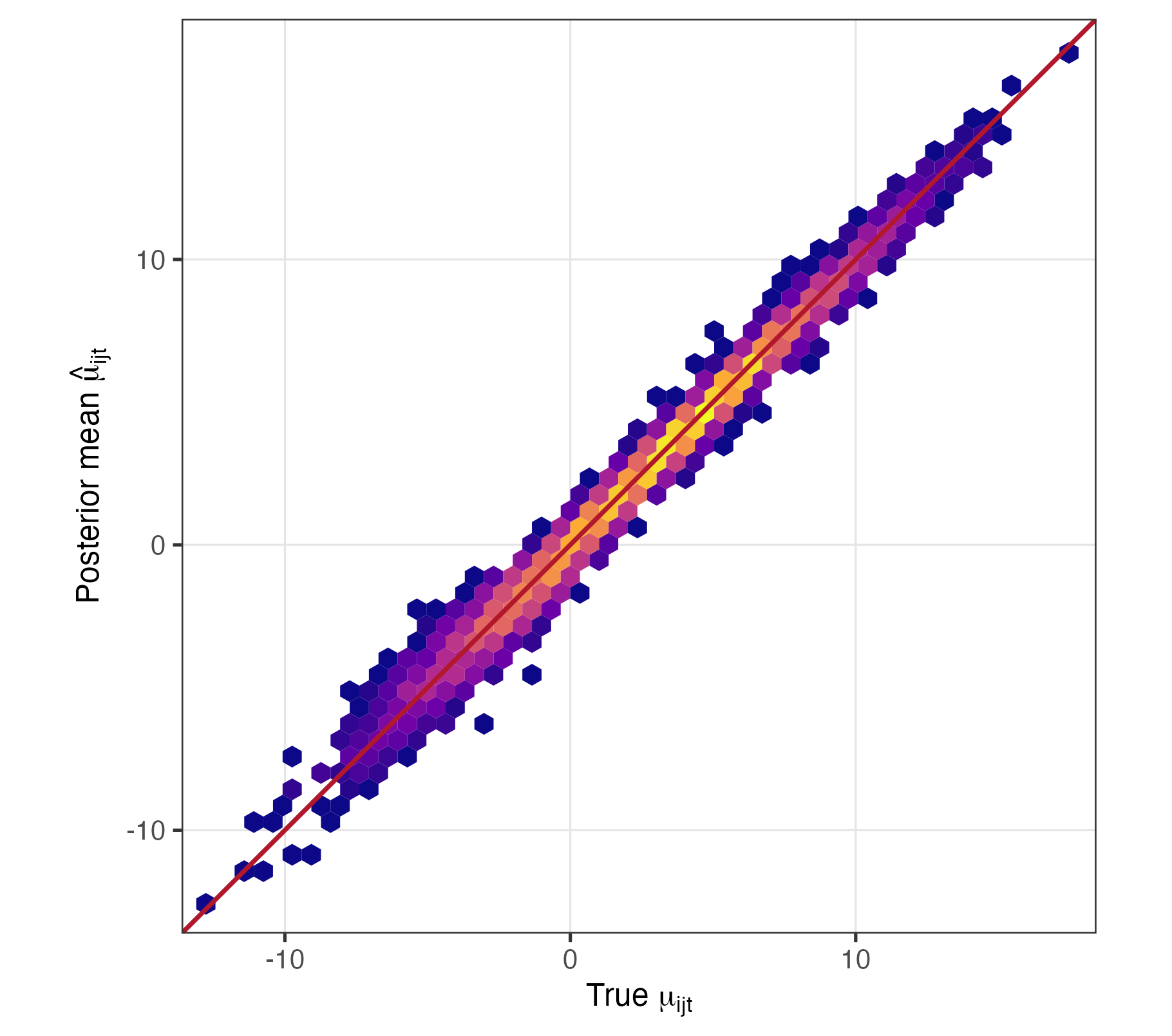} &
  \includegraphics[width=0.46\textwidth]{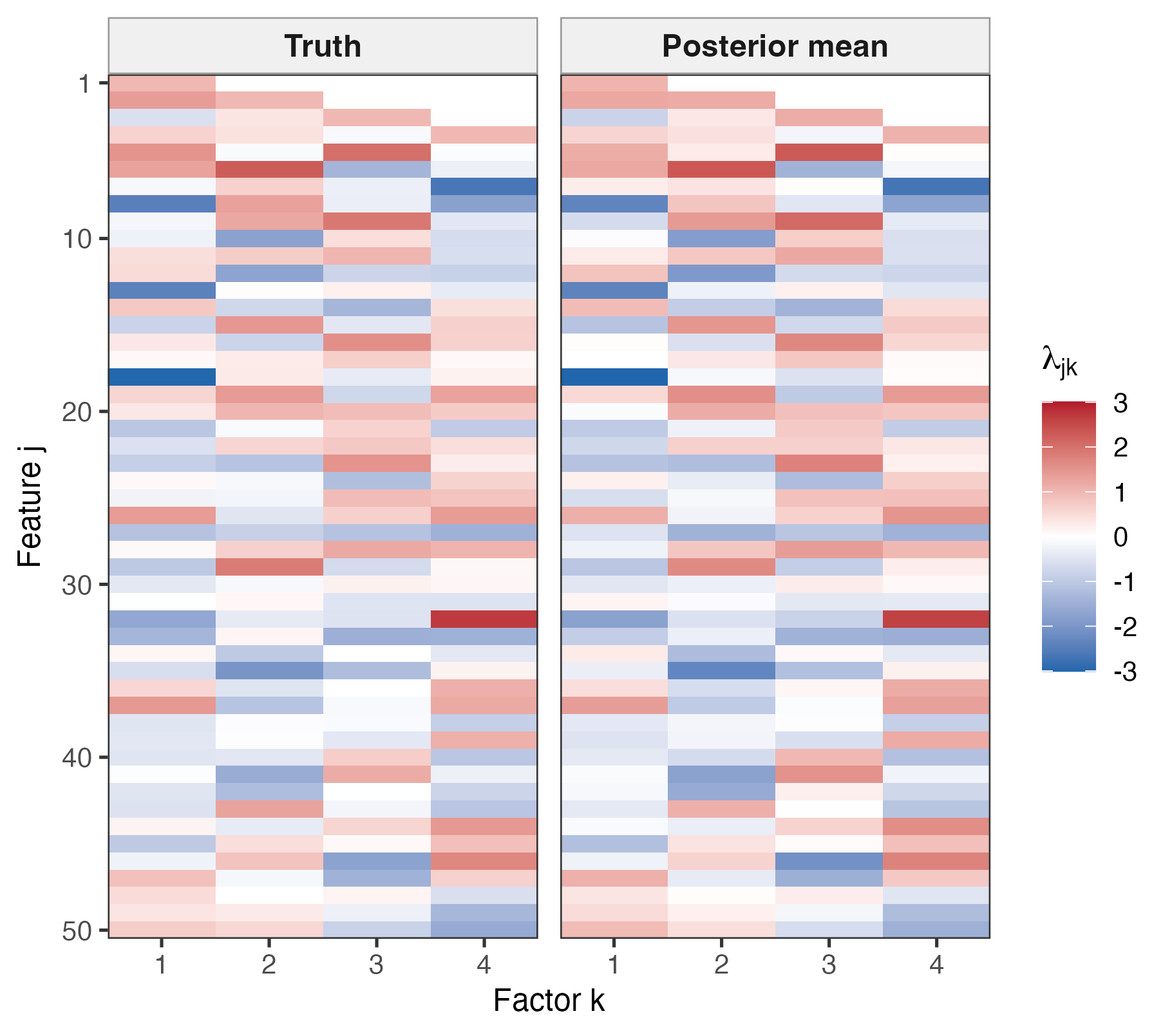} \\
  (a) Mean $\widehat{\mu}_{ijt}$ versus truth $\mu_{ijt}^{\mathrm{tr}}$ &
  (b) Truth $\bLam^{\mathrm{tr}}$ and posterior mean $\widehat{\bLam}$
\end{tabular}
\caption{[Simulation 1] (a) Posterior mean versus truth for all $N\times T\times P=7{,}000$ values of the full latent mean $\mu_{ijt}=s_{it}+\pi_{ijt}+\blam_j^{\T}\beeta_{it}$; warmer colors indicate more overlapping values, and the red line is the identity line. (b) Heatmaps of the true and posterior mean loading matrices with features in rows and factors in columns.}
\label{fig:sim1-parameter-recovery}
\end{figure}

Figure~\ref{fig:sim1-parameter-recovery} illustrates recovery of the full latent mean process and the factor loadings. In Figure~\ref{fig:sim1-parameter-recovery}(a), the posterior mean of the full latent mean $\mu_{ijt}=s_{it}+\pi_{ijt}+\blam_j^{\T}\beeta_{it}$ closely follows its true value across subjects, features, and times. The group-specific AR coefficients are also accurately recovered, with posterior means (SDs) of \(-0.801\ (0.004)\), \(0.910\ (0.004)\), and \(0.694\ (0.006)\), compared with the true values \(-0.8\), \(0.9\), and \(0.7\), respectively. A good estimate of the mean provides a foundation for the covariance estimate. We compare entries of $\bLam$ directly with the truth in Figure~\ref{fig:sim1-parameter-recovery}(b) since both the true and fitted loading matrices obey the same lower-triangular, positive-diagonal constraint. No post-hoc rotation or sign alignment is used. The good estimation of loading leads to a good estimate of covariance, as reported in Table~\ref{tab:sim1}.

LODI substantially outperforms all three comparators and remains stable under moderate over-specification of the factor dimension, as shown in Table~\ref{tab:sim1}. The estimates of covariance for $K=4$, $6$, $8$, and $10$ are similar in terms of their average covariance RMSE. Supplementary Materials Section~\suppsens provides additional mean and loading recovery results for the over-specified factor dimensions. A direct comparison of the true and posterior mean covariance matrices is provided in Supplementary Materials Section~\suppaddsim and shows close recovery of the principal positive and negative covariance patterns. 

\begin{table}[h]
\caption{[Simulation 1] RMSE of $\Sigma_t$ covariance recovery is averaged across time within each replicate and then across 50 independent replicates. Parentheses report its standard error.}
\label{tab:sim1}
\centering
\small
\renewcommand{\arraystretch}{0.95}
\begin{tabular}{@{}lcc@{}}
\toprule
Method & Fitted $K$ & Average RMSE (SE) \\
\midrule
LODI & $4$ (truth) & 0.840 (0.024) \\
LODI & $6$         & 0.853 (0.020) \\
LODI & $8$         & 0.837 (0.020) \\
LODI & $10$        & 0.883 (0.021) \\
\addlinespace[2pt]
MEFISTO                  & 4 & 3.165 (0.036) \\
SP-BGFM (per-time)       & 4 & 3.557 (0.042) \\
SP-BGFM (multi-domain)   & 4 & 3.742 (0.046) \\
\bottomrule
\end{tabular}
\end{table}

In \underline{Simulation 2}, we generate data from a zero-inflated negative binomial distribution. Let $z_{ijt}\sim\Bern(\pi)$ denote structural absence. If $z_{ijt}=1$, then $y_{ijt}=0$; otherwise $y_{ijt}\sim\NB(\vartheta_{ijt},\phi_{\NB}=3)$. Since the negative binomial distribution does not have a covariance parameter, we put the latent factors on the log of the latent mean. $\log\vartheta_{ijt}=c_i+\blam_j^\T\beeta_{it}$ and $c_i\sim\Nor(3,1)$. The factor process has $\rho=0.7$ and identity innovation covariance. Thus, the data-generating likelihood is misspecified relative to LODI. We use $N=20$, $T=5$, $P=50$, $K^{\rm tr}=4$, $G=3$, and test the zero inflation rate $\pi\in\{0.10,0.30,0.50\}$.

\begin{table}[ht]
\caption{[Simulation 2] RMSE of $\Sigma_t$ covariance recovery under increasing zero inflation is averaged across time within each replicate and then across 50 replicates. Parentheses report its standard error.}
\label{tab:sim2}
\centering
\small
\renewcommand{\arraystretch}{0.95}
\begin{tabular}{@{}lccc@{}}
\toprule
Method & $\pi=0.10$ & $\pi=0.30$ & $\pi=0.50$ \\
\midrule
\textbf{LODI}
& \textbf{0.954 (0.029)}
& \textbf{1.205 (0.036)}
& \textbf{1.657 (0.053)} \\
MEFISTO & 2.194 (0.068) & 2.672 (0.067) & 2.974 (0.085) \\
SP-BGFM (per-time) & 2.241 (0.073) & 2.849 (0.083) & 3.290 (0.085) \\
SP-BGFM (multi-domain) & 2.477 (0.082) & 2.926 (0.072) & 3.298 (0.092) \\
\bottomrule
\end{tabular}
\end{table}

As reported in Table~\ref{tab:sim2}, LODI still performs best across the three structural-zero probabilities $\pi$ even under a misspecified model. This advantage over SP-BGFM cannot be attributed to count preprocessing, because both methods use rounded-log-normal augmentation for the raw counts. Instead, the comparison highlights LODI's autoregressive factor process, which allows the sharing of information across time for estimating the evolution of $\bSig_t$. Fitting SP-BGFM separately at each time point discards the information sharing, whereas treating time points as multiple domains does not impose an ordered temporal process. MEFISTO models smooth temporal factors but is fitted to $\log(Y+0.5)$ without a flexible mean process. Supplementary Materials Section~\suppaddsim reports the realized count zero rates and a representative true-versus-estimated covariance comparison. Structural-zero probabilities $\pi=0.10$, $0.30$, and $0.50$ produce mean observed zero rates of 0.202, 0.378, and 0.556, respectively, because the negative-binomial component can also generate zeros. The posterior mean covariance recovers the principal positive and negative dependence patterns under ZINB misspecification, although some covariance magnitudes are moderately attenuated.

\underline{Simulation 3} evaluates a real-data scale simulation under the zero-inflated negative binomial misspecification introduced in Simulation 2. In the real-data analysis of Section~\ref{sec:real-data}, we have 20 subjects measured on 54 OTUs from 3 groups at 7 time points. Thus, we use $N=20$, $T=7$, $P=54$, $K^{\rm tr}=8$, and $G=3$, and generate the latent log mean as $\log\vartheta_{ijt}=c_i+\blam_j^\T\beeta_{it}$. Motivated by the structured empirical OTU dependence shown in Figure~\ref{fig:intro-data-motivation}, we set $K^{\rm tr}=8$ to generate a moderately complex low-rank covariance structure. To reproduce the count scale, we draw $c_i\sim\Nor(\bar{\ell},s_\ell^2)$ using the empirical log-library-size mean $\bar{\ell}\approx10.76$ and empirical standard deviation $s_\ell\approx0.77$ obtained from the real data. We set the structural-zero probability to the empirical zero rate, $\pi=0.189$, and use a factor process with $\rho=0.7$ and identity innovation covariance. We let the negative binomial dispersion parameter equal to $\phi_{\mathrm{NB}}=0.745$, obtained by averaging the valid feature-specific method-of-moments estimates from the positive
real-data counts.

\begin{figure}[h]
\centering
\begin{tabular}{@{}cc@{}}
  \includegraphics[width=0.47\textwidth]{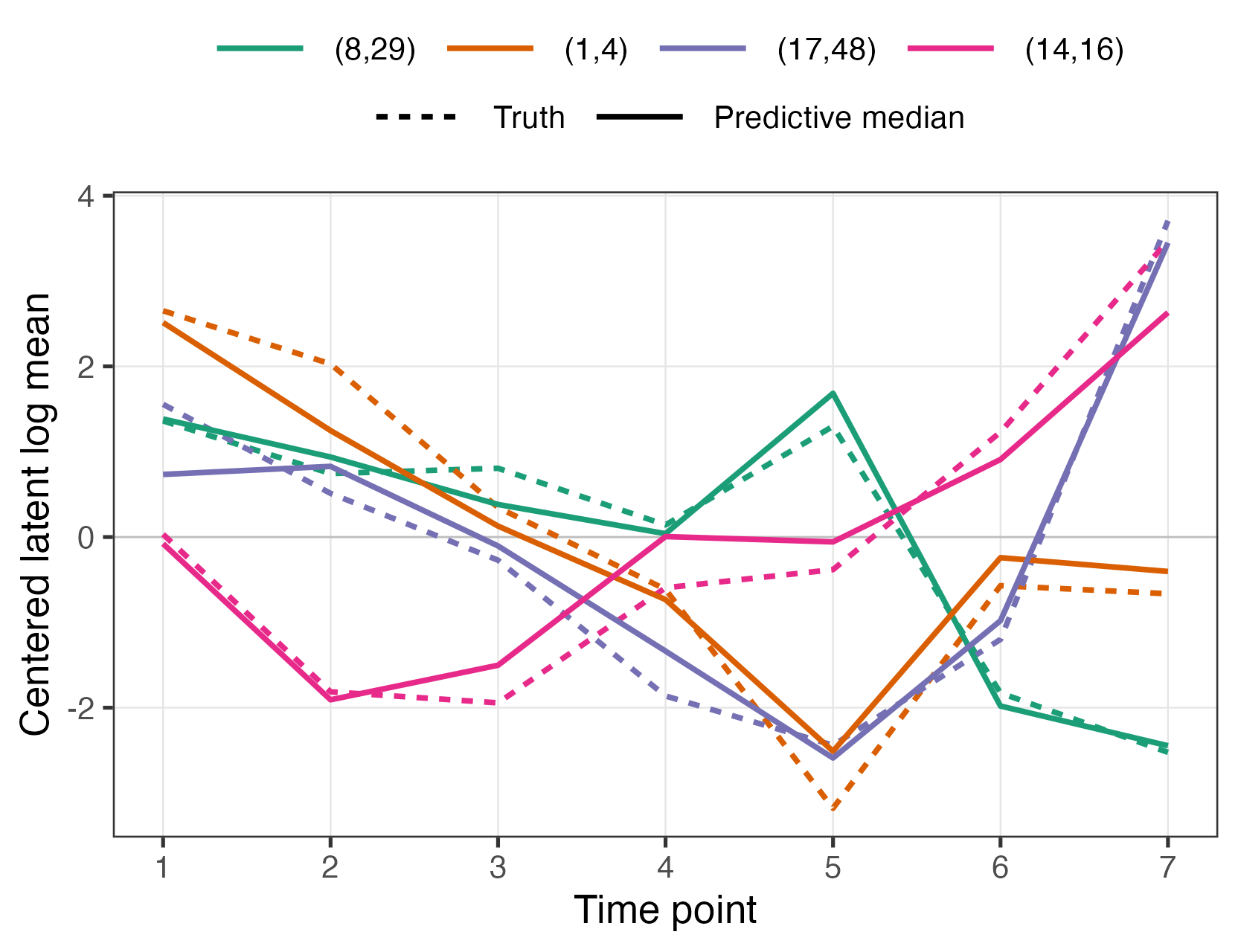} &
  \includegraphics[width=0.47\textwidth]{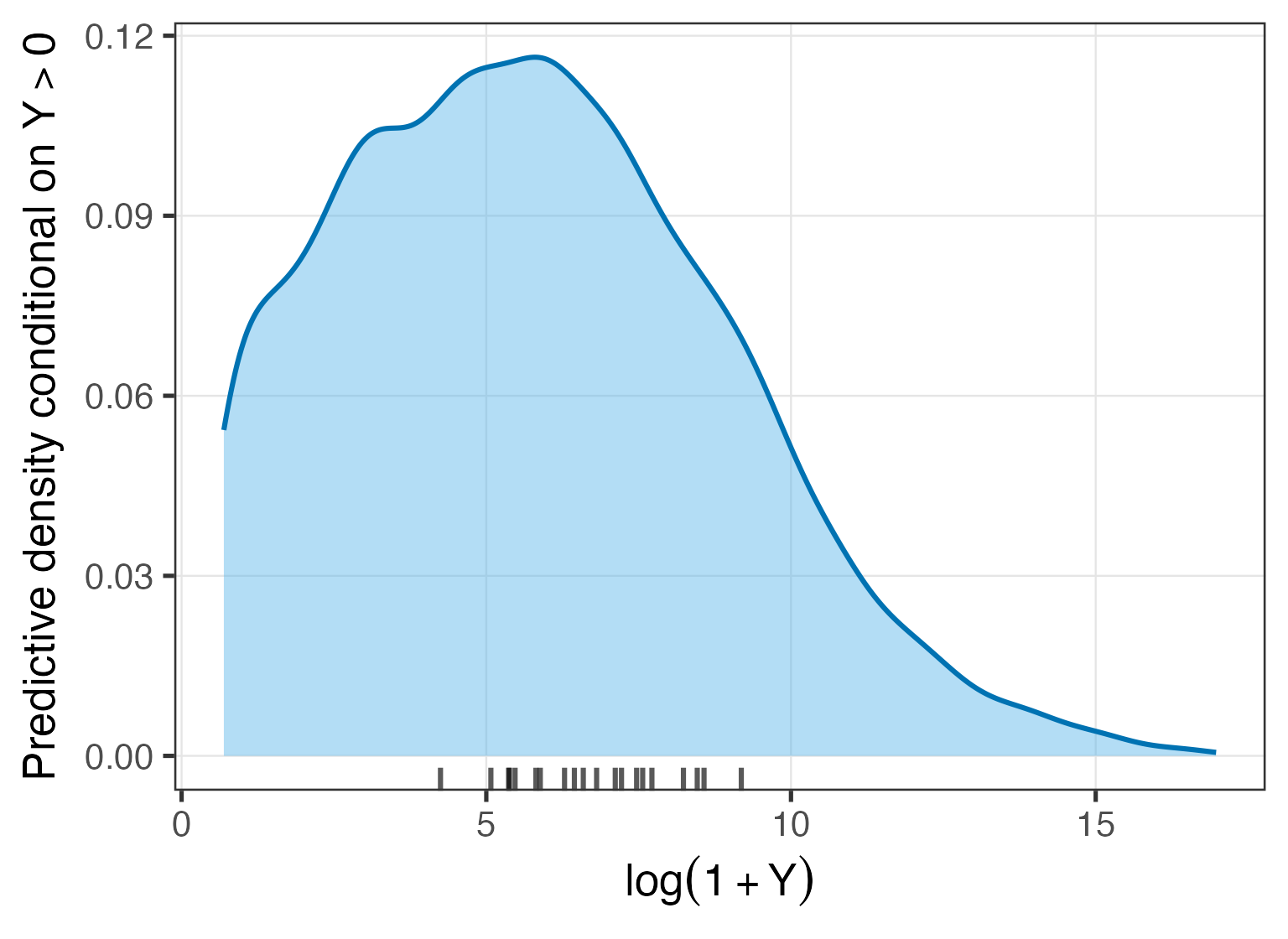} \\
  (a) Four subject--OTU trajectories & (b) OTU 36 at $t=1$ \\[4pt]
  \includegraphics[width=0.47\textwidth]{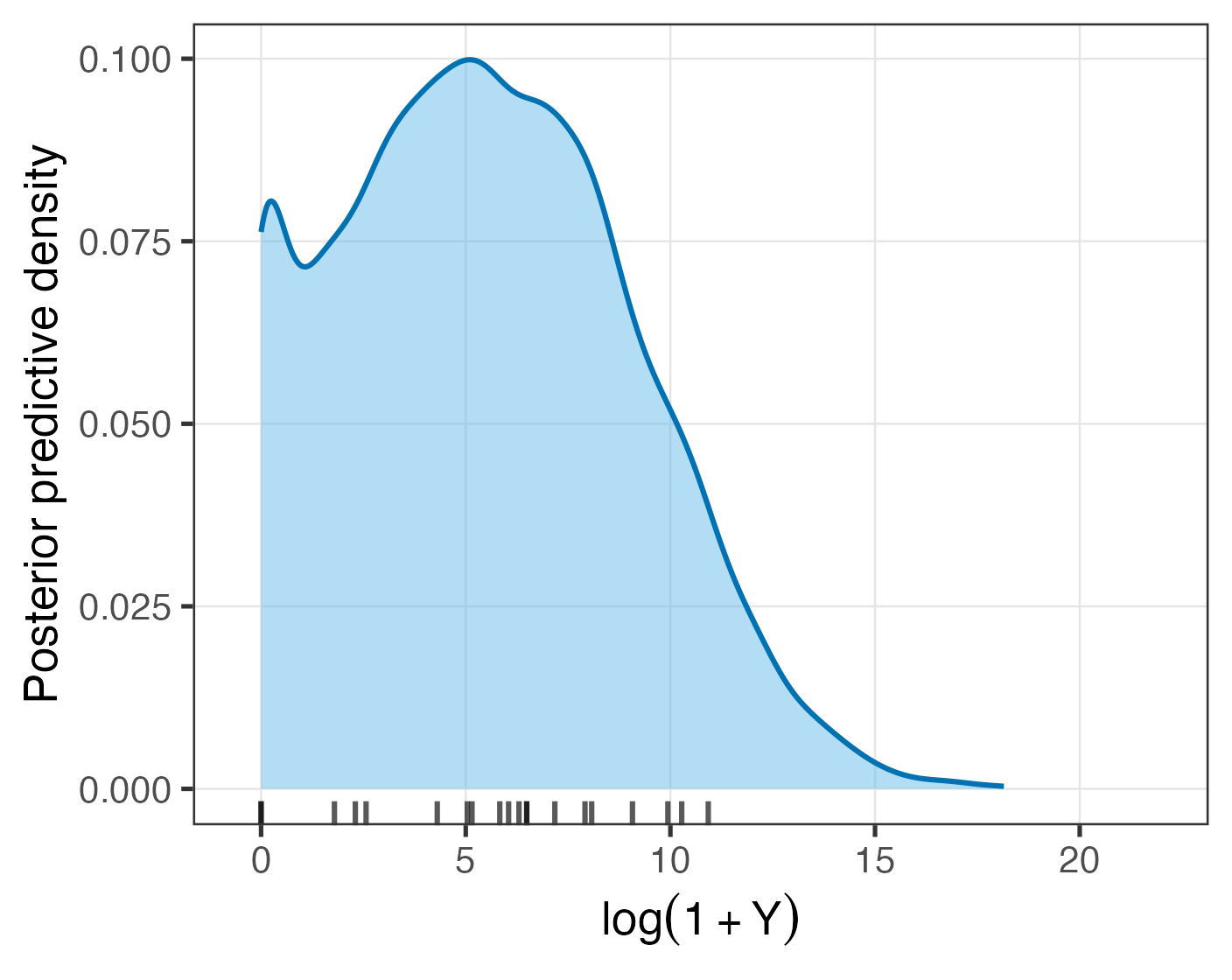} &
  \includegraphics[width=0.47\textwidth]{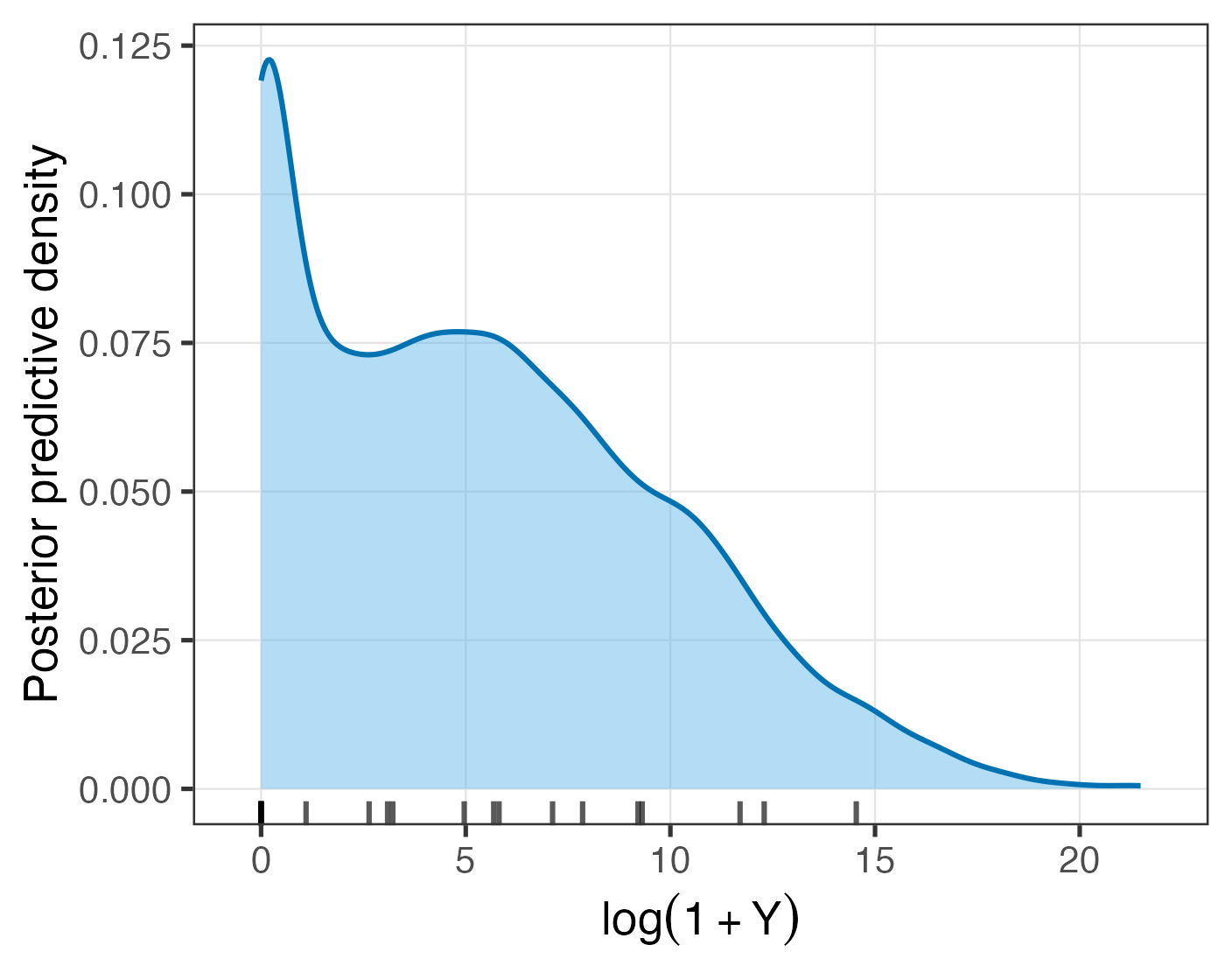} \\
  (c) OTU 7 at $t=7$ & (d) OTU 53 at $t=6$
\end{tabular}
\caption{[Simulation 3] Panel (a) compares true normalized latent log-means (dashed) with posterior mean estimates (solid) for $(i,j)=(8,29),(1,4),(17,48)$, and $(14,16)$. Panels (b)--(d) show posterior predictive densities of $\log(1+Y_{ijt}^{\rm rep})$ for OTU 36 at $t=1$, OTU 7 at $t=7$, and OTU 53 at $t=6$, respectively. Black rug marks denote the 20 observed counts.}
\label{fig:sim3-mean-ppc}
\end{figure}

Figure~\ref{fig:sim3-mean-ppc} examines temporal mean trajectories and posterior predictive check for goodness of fit. Panel (a) shows posterior mean trajectories for four subject–OTU pairs (solid), tracking closely the true normalized latent log-mean processes (dashed) with distinct temporal shapes. For the posterior predictive check, we compare summaries of posterior predictive samples with the corresponding summaries of the observed or simulated counts. Panels (b)--(d) provide distributional checks across increasing sparsity. Panel (b) evaluates the positive-count distribution in a setting with no observed zeros, and Panels (c) and (d) represent increasingly sparse settings. As the zero rate in the observed data increases, the posterior predictive distribution adaptively places more mass near zero. This pattern indicates that our nonparametric approach reproduces zero-rich count behavior at any time point without a separate zero-inflation indicator.

Table~\ref{tab:sim3} evaluates covariance recovery for LODI and comparators across all replicates. LODI has the smallest average RMSE. The absolute RMSE values are not directly comparable with those in Simulation 2 because Simulation 3 uses different dimensions and a distinct real-data-calibrated data-generating setting. Representative post-burn-in trace plots for the global temporal parameters and selected covariance entries are provided in Supplementary Materials Section~\suppaddsim; the traces fluctuate around stable levels without visible long-term drift.

\begin{table}[ht]
\caption{[Simulation 3] RMSE of $\Sigma_t$ covariance recovery is averaged across time within each replicate and then across 50 replicates, and parentheses report standard errors.}
\label{tab:sim3}
\centering
\small
\renewcommand{\arraystretch}{0.95}
\begin{tabular}{@{}lc@{}}
\toprule
Method & Average RMSE (SE) \\
\midrule
\textbf{LODI} & \textbf{2.002 (0.039)} \\
MEFISTO & 2.376 (0.057) \\
SP-BGFM (per-time) & 2.284 (0.074) \\
SP-BGFM (multi-domain) & 2.976 (0.083) \\
\bottomrule
\end{tabular}
\end{table}

\section{Real data analysis}\label{sec:real-data}

We apply LODI to a longitudinal mouse-gut microbiome dataset comprising $N=20$ mice measured at $T=7$ time points across $P=54$ OTUs. The mice belong to three genotype groups: ACK/G ($n=4$), HACK/G ($n=10$) harboring $Apc^{S580/+}$ and $Kras^{G12D/+}$ mutations, with/without two alleles of $AhR$ ($AhR^{f/f}$, abbreviated as H), and WT (wild type, no tumor control; $n=6$), measured at weeks 0, 1, 3, 7, 11, 15, and 19 \citep{han2021loss}. At W15, tumor severity is significantly different across genotypes: WT $<$ ACK/G $<$ HACK/G.

Genomic DNA is extracted from flash-frozen fecal samples according to the DNeasy PowerSoil kit (Qiagen) instructions using an automated QIAcube (Qiagen), as previously described by \citet{yang2020effect}. DNA quantity is assessed using the 260/280 absorbance ratio measured on a NanoDrop spectrophotometer (Thermo Fisher Scientific); all samples have a ratio of approximately 1.8. The V4 region of the bacterial 16S rRNA gene is sequenced using $2\times250$ bp paired-end reads on a MiSeq platform (Illumina). The reads are processed using QIIME~2 (v2024.2) \citep{bolyen2019reproducible}. Demultiplexed paired-end sequences are imported into QIIME~2 and denoised using DADA2 \citep{callahan2016dada2}, with 20 base pairs trimmed from both ends. The raw count matrix has a total observation count summing to $\approx 1.6 \times 10^7$, with library sizes per (subject, time) ranging from $1.0\times 10^4$ to $3.1 \times 10^5$. The overall zero rate is $18.9\%$, but per-feature zero rates are highly heterogeneous (0\% to 92.1\%). We fit LODI with $K=10$ latent factors (given that overspecification is benign as shown in simulations) and 400{,}000 MCMC iterations. 

\begin{figure}[ht]
    \centering
    \includegraphics[width=0.95\linewidth]{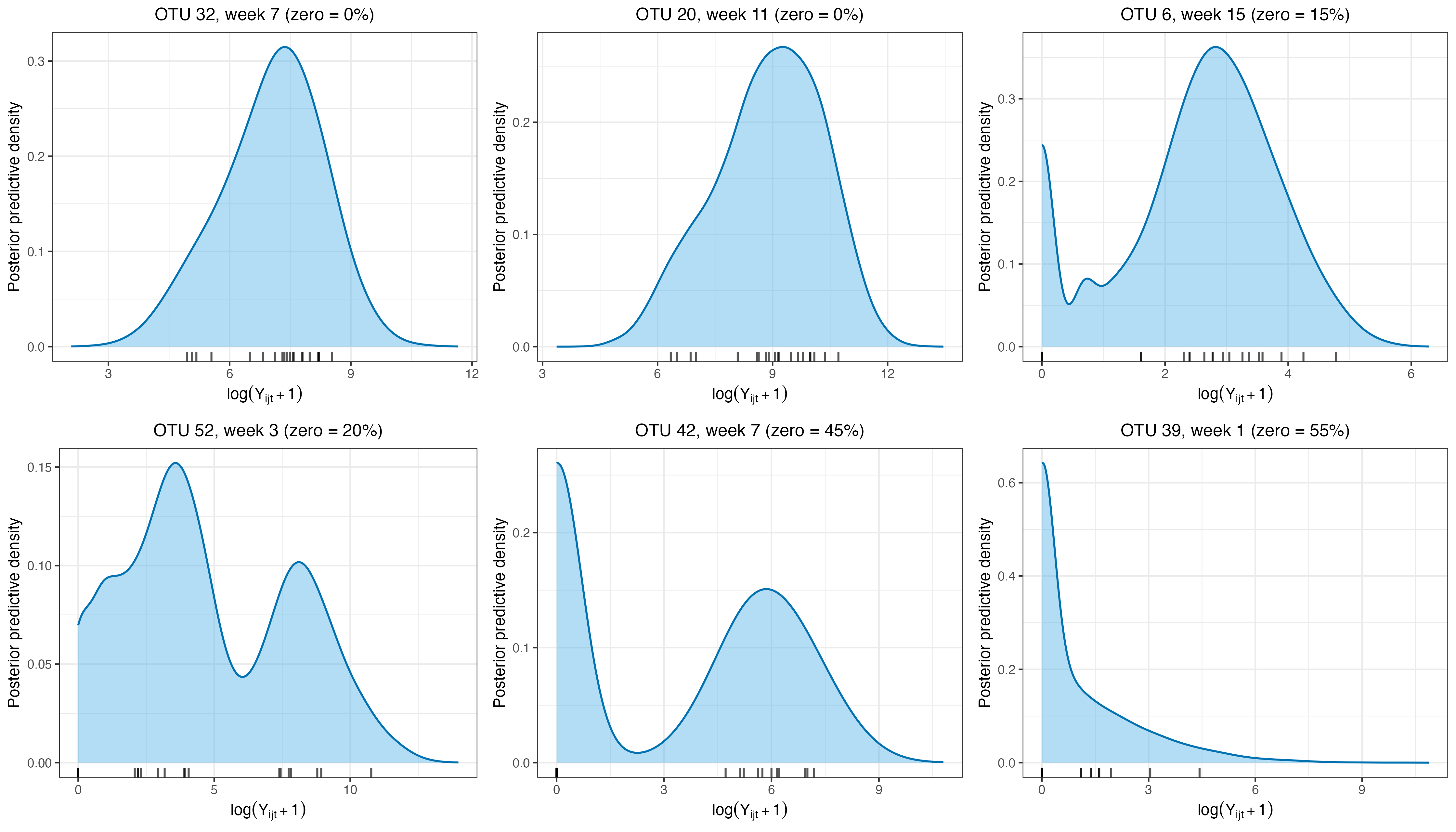}
    \caption{[Real Data] Posterior predictive densities (blue) of $\log(Y_{ijt}^{\rm rep}+1)$ for six representative OTU--time
    pairs, ordered by increasing time-specific observed zero rate.
    Each panel selects one OTU and one week and pools predictions across
    the 20 mice; black rug marks show the corresponding 20 observed
    counts. OTU 52 at week 3 illustrates multimodal flexibility, OTU 6
    at week 15 represents moderate sparsity, and OTUs 42 and 39
    illustrate recovery under substantial sparsity.}
    \label{fig:ppc-per-otu}
\end{figure}

\begin{figure}[ht]
\centering
\begin{tabular}{cc}
  \includegraphics[width=0.47\textwidth]{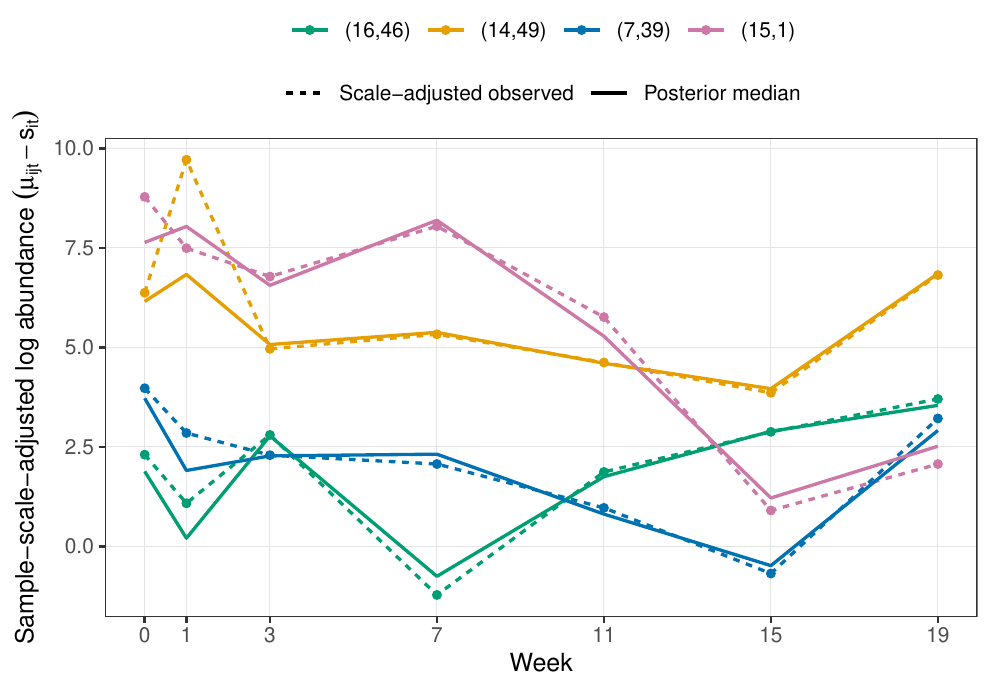} &
  \includegraphics[width=0.47\textwidth]{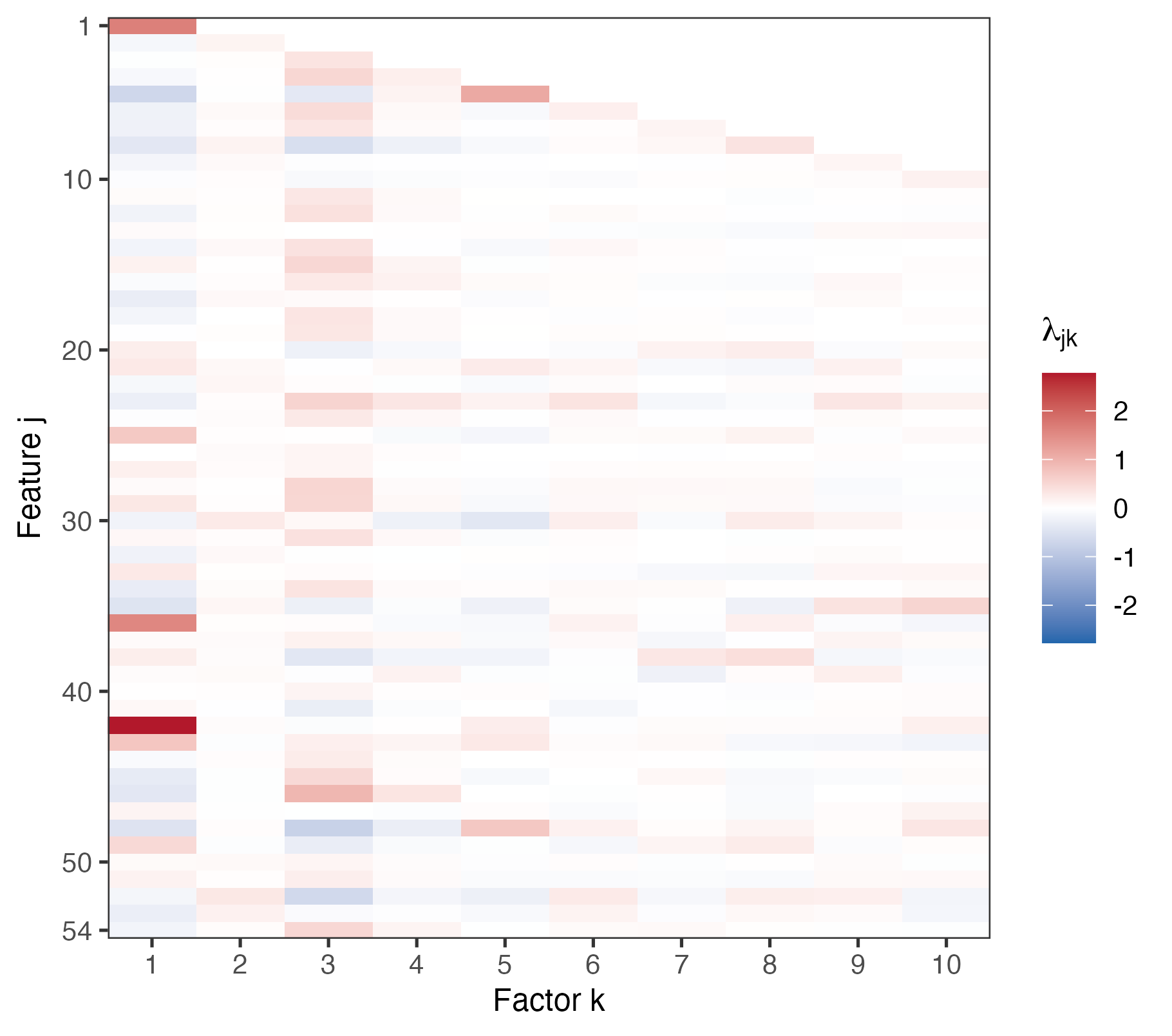} \\
  (a) Sample-scale normalized process $\mu_{ijt}-s_{it}$ & (b) Posterior mean factor loadings $\widehat{\bLam}$
\end{tabular}
    \caption{[Real Data] Panel (a) compares posterior median trajectories of $\mu_{ijt}-s_{it}=\pi_{ijt}+\blam_j^{\T}\beeta_{it}$ (solid) with $\log(Y_{ijt}+0.5)$ minus the posterior median of $s_{it}$ (dashed, with points), for four illustrative subject--OTU pairs indexed as $(i,j)$. The observed adjustment uses the posterior sample-scale estimates. Panel (b) shows the posterior mean of the identified loading matrix $\widehat{\bLam}$.}
    \label{fig:realdata-trajectories}
\end{figure}

To assess goodness of fit, we conduct posterior predictive checks and compare the fitted abundance process to the observed one. Figure~\ref{fig:ppc-per-otu} shows the posterior predictive density of $\log(Y_{ijt}^{\rm rep}+1)$ for six representative OTU--time pairs ordered by their time-specific observed zero rates. LODI's posterior predictive distribution flexibly captures the shape of the count distribution for abundant OTUs with few zeros (e.g., OTU 32 at Week 7, zero rate 0\%), bimodal OTUs (OTU 52 at Week 3, zero rate 20\%), and sparse OTUs (e.g., OTU 39 at Week 1, zero rate 55\%). This flexibility benefits directly from LODI's DP-mixture prior on $\pi_{ij1}$. Figure~\ref{fig:realdata-trajectories}(a) compares posterior median trajectories of the sample-scale normalized process $\mu_{ijt}-s_{it}$ with observed log counts normalized by the fitted sample-scale estimate, for four illustrative subject--OTU pairs.
This fitted decomposition diagnostic complements the posterior predictive checks. The feature-specific $\pi_{ijt}$ trajectories are provided in Supplementary Materials Section~\suppaddreal. The posterior means of $\phi_g$, which capture the feature-specific abundance process $\pi_{ijt}$, are 0.993, 0.994, and 0.995 across the three genotype groups, and the innovation variance is $V=0.473~(\pm0.040)$. The estimated innovation variance allows departures from the preceding abundance level. The shared factor process has posterior mean $\rho=0.347~(\pm0.065)$ and 95\% credible interval $[0.221,0.474]$, indicating substantially weaker temporal persistence than the feature-specific abundance process.

\begin{figure}[ht]
    \centering
    \includegraphics[width=0.95\linewidth]{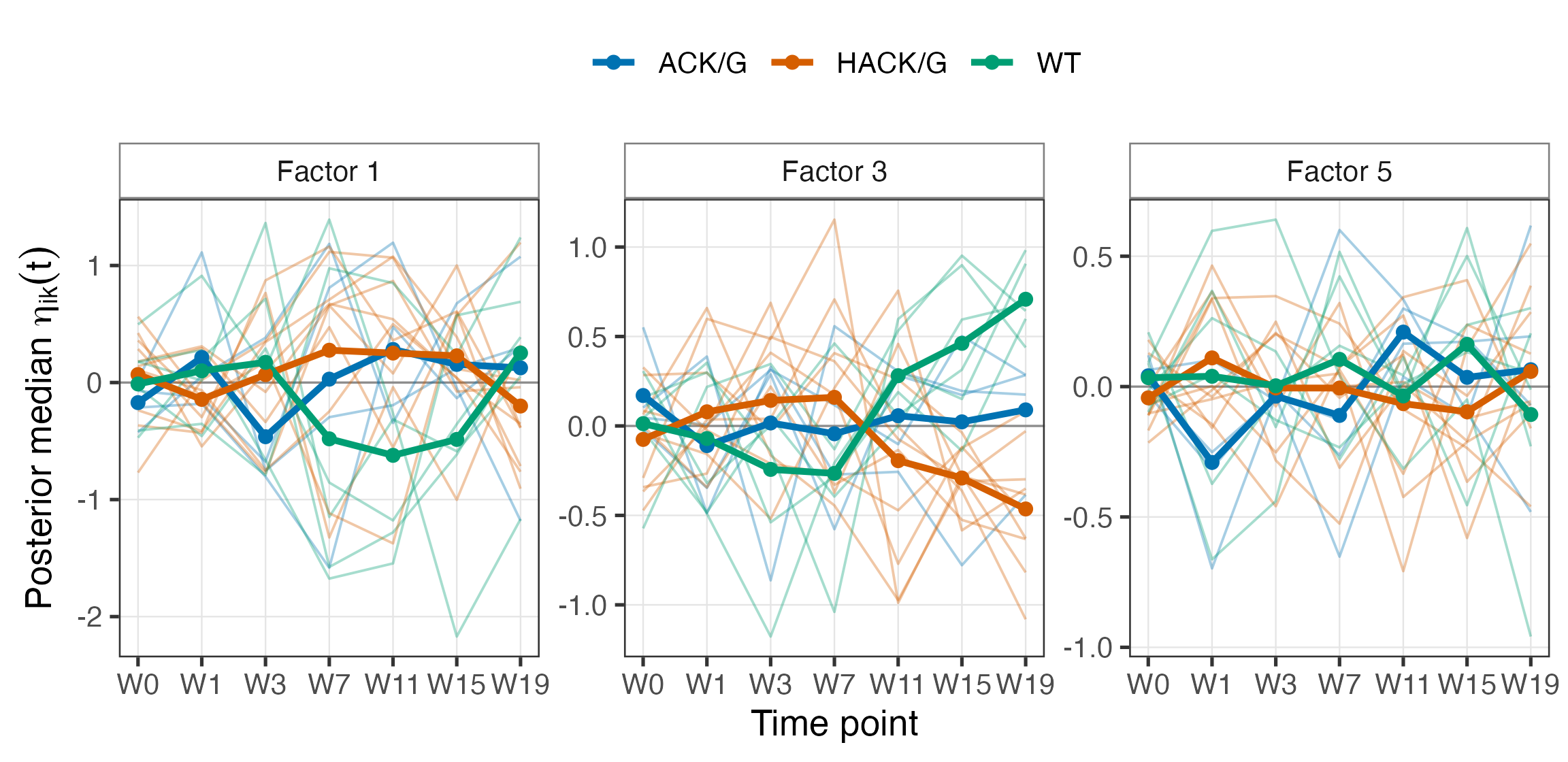}
    \caption{[Real Data] Posterior mean factor-score trajectories $\eta_{ikt}$ for the three dominant loading directions. Thin curves show individual subjects and thick curves show genotype-group averages. Factor scores are centered across subjects at each time point.}
    \label{fig:realdata-eta}
\end{figure}

Figure~\ref{fig:realdata-trajectories}(b) displays the posterior mean of the identified loading matrix. Its effective rank is substantially smaller than the prespecified factor dimension $K=10$. Measured by $\|\widehat{\blam}_{\cdot k}\|^2/\|\widehat{\bLam}\|_F^2$, factors 1, 3, and 5 contribute 56.4\%, 21.6\%, and 8.2\% of the total magnitude, respectively, whereas each remaining factor contributes only 1.2\%--2.7\%. Motivated by the three dominant loading columns in Figure~\ref{fig:realdata-trajectories}(b), we plot posterior mean trajectories of $\eta_{ikt}$ for factors 1, 3, and 5 in Figure~\ref{fig:realdata-eta}. Thin curves represent individual subjects, and thick curves represent genotype-group averages. Factor 1 exhibits the strongest short-term reversals, including a decline from W1 to W3 followed by recovery at W7 for the ACK/G group and a decline at W7 followed by recovery at W19 for the WT group. Factor 3 shows increasing separation after W7: the WT group moves upward, reaching 0.71 at W19, whereas HACK/G moves downward to $-0.46$. Factor 5 remains closer to zero at the group level, although individual subjects still fluctuate.

The magnitude of $\lambda_{jk}$ measures the strength of OTU $j$'s association with factor $k$. Its sign determines whether the factor-specific contribution $\lambda_{jk}\eta_{ikt}$ increases or decreases as the factor score increases. At W15, the posterior genotype means for factor 1 are $-0.491$, $0.164$, and $0.229$ for WT, ACK/G, and HACK/G, respectively, following the reported tumor-severity ordering. Positive factor-1 loadings therefore associate this loading direction with higher latent abundance in the more severe genotype groups. The largest positive posterior mean loadings occur for OTU 42 (family Clostridiaceae; 2.77), OTU 1 (genus \textit{Clostridium}, family Clostridiaceae; 1.65), and OTU 36 (order RF39; 1.56). For factor 3, the W15 posterior genotype means are $0.426$, $0.019$, and $-0.263$, respectively. Negative loadings consequently give factor-specific contributions in the WT $<$ ACK/G $<$ HACK/G direction, whereas positive loadings give contributions in the opposite direction. The largest negative posterior mean loadings occur for OTU 48 (genus \textit{Enterococcus}; $-0.77$), OTU 52 (genus \textit{Akkermansia}; $-0.64$), and OTU 8 (genus \textit{Sutterella}; $-0.54$); the largest positive loading occurs for OTU 46 (genus \textit{Butyricicoccus}; 0.93). 
The factor-3 contribution of \textit{Butyricicoccus} in the direction opposite to genotype severity is biologically compatible with lower abundance in advanced colorectal cancer and less aggressive tumors following \textit{B. pullicaecorum} administration in mice \citep{huang2020gut}. Colorectal cancer studies also report enrichment of \textit{Enterococcus}-related taxa \citep{wang2012structural}, although our loading-based contribution does not establish a monotonic genus-wide abundance pattern. 

\begin{figure}[ht]
    \centering
    \begin{tabular}{cc}
      \includegraphics[width=0.47\textwidth]{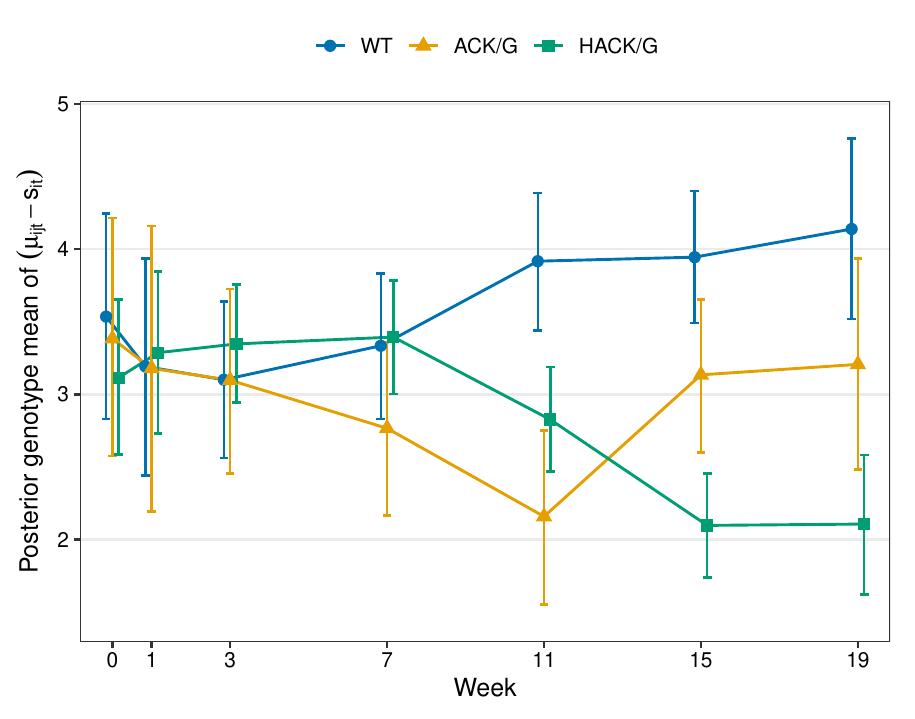} &
      \includegraphics[width=0.47\textwidth]{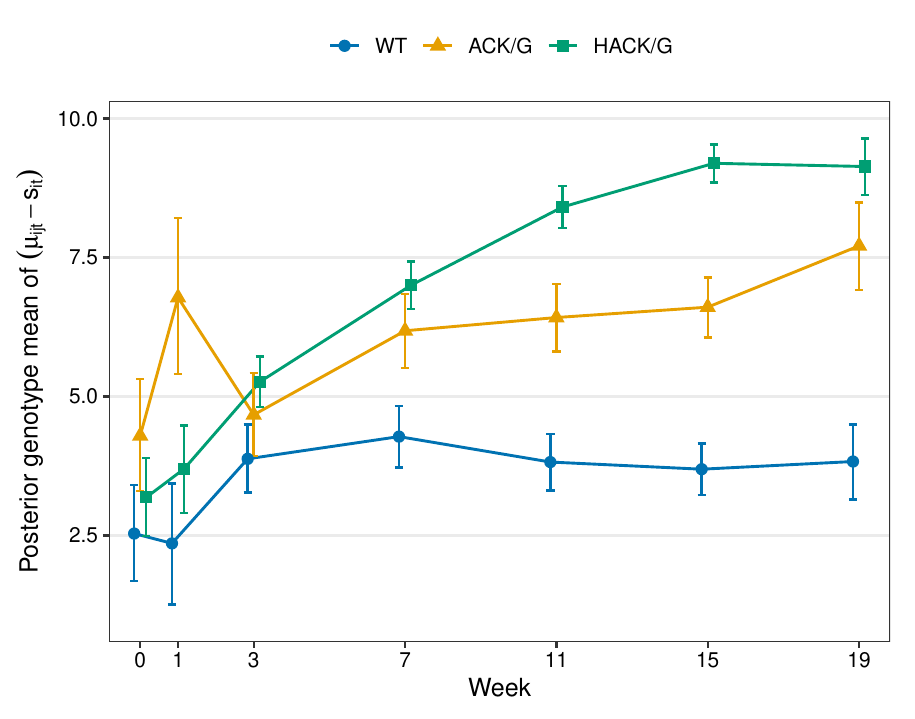} \\
      (a) OTU 14 (Ruminococcaceae--\textit{Clostridium}) &
      (b) OTU 52 (\textit{Akkermansia})
    \end{tabular}
    \caption{[Real Data] Longitudinal posterior sample-scale normalized abundance processes for (a) OTU 14 (\textit{Clostridium}, family Ruminococcaceae) and (b) OTU 52 (\textit{Akkermansia}). Points and lines show posterior genotype means of $\mu_{ijt}-s_{it}$, and error bars show pointwise 95\% credible intervals. The process is averaged across mice within each genotype on its original latent log scale. }
    \label{fig:realdata-genotype-abundance}
\end{figure}

To examine genotype differences over time, we calculate genotype-specific means of the sample-scale normalized mean process $\mu_{ijt}-s_{it}$ in Figure~\ref{fig:realdata-genotype-abundance}. At each posterior draw, this process is averaged across mice within genotype on its original log scale. 
Among the taxa, we focus on OTU 14 (\textit{Clostridium}, family Ruminococcaceae) and OTU 52 (\textit{Akkermansia}), for which all three pairwise genotype differences have 95\% credible intervals excluding zero at W15. For OTU 14, early genotype differences through W7 remain uncertain. At W11, WT exceeds both mutant groups, and at W15 all three differences follow WT $>$ ACK/G $>$ HACK/G, with posterior genotype means of 3.945, 3.134, and 2.098, respectively. At W19, HACK/G remains significantly lower than both other groups. The lower estimated abundance in the more severe genotype groups is directionally compatible with depletion of some butyrate-producing clostridial taxa in human colorectal cancer \citep{wang2012structural}. For OTU 52, all three pairwise differences support the abundance ordering WT $<$ ACK/G $<$ HACK/G at W7, W11, W15, and W19, matching the reported tumor-severity ordering. At W15, the posterior genotype means are 3.690, 6.606, and 9.195, respectively. The HACK/G trajectory rises through W15 and remains elevated at W19, whereas the WT mean changes relatively little after W3. This increase differs from the depletion and protective effects reported for \textit{Akkermansia muciniphila} in some colorectal cancer settings \citep{zhang2023akkermansia}, while other mouse experiments report increased tumor development following its administration \citep{wang2022akkermansia}. 

We also report the posterior correlation estimates over time and the strongly associated OTU pairs in Supplementary Materials Section~\suppaddreal. Several estimated associations are consistent with previous gut-microbiome studies. In particular, the positive association between OTU 20 (genus \textit{Allobaculum}) and OTU 49 (genus \textit{Bifidobacterium}) ranges from 0.303 to 0.780 across weeks. This agrees with the significant positive co-abundance relationship reported by \citet{wang2019core} across 101 healthy mice and with the Spearman correlation of 0.78 reported by \citet{tachon2013intestinal} in aged mice. The positive association between OTU 20 and OTU 38 (family Coriobacteriaceae), with posterior mean correlations ranging from 0.318 to 0.706, is also consistent at a broader taxonomic level with the positive association between Erysipelotrichaceae---the family containing \textit{Allobaculum}---and Coriobacteriaceae reported by \citet{liu2022gut}. Finally, the consistently positive association between OTU 5 (family Enterobacteriaceae) and OTU 48 (genus \textit{Enterococcus}), with posterior mean correlations ranging from 0.341 to 0.483, is concordant with the identification of \textit{Enterococcus} species as prominent co-colonizers of Enterobacteriaceae in a large-scale analysis of human gut metagenomes \citep{yin2025ecological}. These agreements support the biological plausibility of the estimated dependence structure, although the correlations do not necessarily represent direct microbial interactions and may partly reflect shared host or environmental effects.

\FloatBarrier
\section{Discussion}\label{sec:discussion}

We have introduced LODI, a Bayesian longitudinal factor model for microbiome counts. It combines a rounded log-normal latent representation of count data, a DP-mixture baseline, a group-specific AR(1) abundance process $\pi_{ijt}$, and a shared AR(1) factor process $\beeta_{it}$.  LODI is capable of capturing both the temporal
evolution of microbial abundances and dynamic cross-feature dependence.
Posterior inference uses FFBS updates for both longitudinal processes within an Rcpp-based Gibbs sampler. Simulations under both the assumed model and misspecified zero-inflated negative binomial mechanisms demonstrate improved covariance recovery over existing methods, and the mouse-gut microbiome temporal analysis illustrates how LODI separates persistent subject-feature abundance trajectories from shared variation across OTUs.

Several extensions could further increase flexibility while preserving this separation. The current model uses a time-invariant loading matrix $\bLam$, which pools information across time and makes the factor directions directly comparable. A dynamic loading process $\bLam_t$ could permit the OTU-specific dependence pattern itself to evolve, rather than allowing temporal covariance changes only through the factor process and residual variance. Because an unrestricted $P\times K$ loading matrix at every time point is hard to estimate when $N\ll P$, a practical version would require temporal smoothing, shrinkage toward a common loading matrix, and identifying constraints that remain coherent over time.

The baseline distribution could also vary across biological groups or covariates. A hierarchical Dirichlet process could define genotype- or OTU-specific mixing distributions that share atoms but use different weights \citep{teh2006hierarchical}. Alternatively, a dependent Dirichlet process could allow the entire baseline distribution to change flexibly with genotype or other subject-level covariates \citep{quintana2022dependent}. These extensions may reveal group-specific baseline heterogeneity, although reliable uncertainty quantification would require more subjects within each group.

\begin{samepage}
\section*{Acknowledgments}
Robert S. Chapkin and Yang Ni are also affiliated with the Texas A\&M Regional Center of Excellence in Cancer. Correspondence should be addressed to Shuangjie Zhang
(\href{mailto:shuangjie.zhang@austin.utexas.edu}{shuangjie.zhang@austin.utexas.edu}).
\end{samepage}

\section*{Funding}
This work was supported in part by National Institutes of Health Grants R01-GM148974, R01-GM163238, R35-CA197707, R01-ES025713, and R01-CA202697, CPRIT Grant RP230204, and the Allen Endowed Chair in Nutrition \& Chronic Disease Prevention.

\bibliographystyle{plainnat}
\bibliography{references}
\end{document}